\documentclass{svjour3}  
\smartqed  

\usepackage{amsmath}
\usepackage{graphicx}
\usepackage{color}
\usepackage{graphicx}

\usepackage[colorlinks=true,linkcolor=blue,citecolor=blue]{hyperref}

\newcommand{\be}{\begin{equation}}
\newcommand{\ee}{\end{equation}}
\newcommand{\bea}{\begin{eqnarray}}
\newcommand{\eea}{\end{eqnarray}}

\newcommand{\eref}[1]{Eq.~(\ref{#1})}%
\newcommand{\fref}[1]{Fig.~\ref{#1}} %
\newcommand{\sref}[1]{Sec.~\ref{#1}}%

\begin{document}

\title{Splash in an inhomogeneous gas in one dimension: Exact analysis and molecular dynamics simulations}      
\titlerunning{Blast waves}
\author{Amit Kumar \and R. Rajesh}

\institute{Amit Kumar \at
              	The Institute of Mathematical Sciences, CIT Campus, Taramani, Chennai 600113, India \\
              	Homi Bhabha National Institute, Training School Complex, Anushakti Nagar, Mumbai 400094, India\\
               	\email{kamit@imsc.res.in}  	    
         \and
             R. Rajesh \at
              	The Institute of Mathematical Sciences, CIT Campus, Taramani, Chennai 600113, India \\
              	Homi Bhabha National Institute, Training School Complex, Anushakti Nagar, Mumbai 400094, India\\
               	\email{rrajesh@imsc.res.in}           
}

\date{Received: \today / Accepted: }

\maketitle

\begin{abstract}
We investigate the splash phenomenon resulting from the energy input at the interface between a vacuum and an inhomogeneous gas with density profile $\rho(r) = \rho_0 r^{-\beta}$. The energy input causes the formation of ballistic spatters that propagate into the vacuum, leading to a decay of the total energy in the inhomogeneous medium following a power law, $E(t) \sim t^{-\delta_s}$. We determine exactly the exponents $\delta_s$ by solving the Euler equation using a self-similar solution of second kind for different values of $\beta$. These exponents are further validated through event-driven molecular dynamics simulations. The determination of these exponents also allows us to numerically determine the spatio-temporal dependence of the density, velocity and temperature.
\end{abstract}

\section{\label{sec1-Introduction}Introduction}
The shock propagation emanating from a point explosion is a classic problem in gas dynamics~\cite{landaubook,whitham2011linear,barenblatt1996scaling,sedov_book}. The shock grows radially outwards in time with a shock front separating the moving gas from the ambient gas. The systems, after initial transients, Initially,  goes into the hydrodynamic regime, where main mode of energy transport  is the motion of the particles. In the hydrodynamic regime, the system is described in terms of local fields of thermodynamic quantities: density, velocity, temperature, and the pressure. These thermodynamic quantities are single valued and continuous within the disturbed region, but across the shock front they change abruptly. The amplitude of this abrupt change is described by the Rankine-Hugoniot boundary conditions~\cite{landaubook,whitham2011linear,sedov_book}. The spatio-temporal behavior of  density, $\rho(r,t)$, radial velocity, $u(r,t)$, temperature, $T(r,t)$, and pressure $p(r,t)$ has been studied in homogeneous as well in inhomogeneous medium~\cite{whitham2011linear,barenblatt1996scaling,sedov_book}. 

The gas could be initially spatially isotropic or anisotropic. We will refer to the isotropic case as the blast problem, which is very well studied for both the homogeneous and inhomogeneous gas where $\rho(r,0) = \rho_0 r^{-\beta}$. Due to point symmetry, these thermodynamic quantities depend only on radial distance $r$ and time $t$. The scaling exponents are easily obtained using dimensional analysis. It is known that near the shock front, the shock is well described by Euler equation, while near the shock center, there is a crossover to a region where heat conduction become important, and the hydrodynamics is  described by the Navier-Stokes equation~\cite{taylor1950formation,taylor1950formation2,jvneumann1963cw,sedov1946,jabeen2010universal,antal2008exciting,joy2021shock3d,joy2021shock,chakraborti2021blast,ganapa2021blast,kumar2022blast,kumar2024shock,kumar2024shock2,singh2023blast,dokuchaev2002self}.

The gas could be initially spatially anisotropic. A special case is one when there is vacuum in half the space and a gas in the other half. Energy is input at a point in the interface. We will refer to this problem as the splash problem. This problem is less studied in comparison to the blast problem. In a recent paper, the splash problem was studied in one dimension when the gas is homogeneous~\cite{chakraborti2022splash}. The energy is reflected back into the vacuum, and in the scaling regime the total energy of  medium (gas) decays as $E(t)\sim t^{-\delta_s}$. The value of $\delta_s= 0.11614383675...$ has been obtained for the homogeneous medium in one dimension by analysing the Euler equation, using self-similarity of the second kind. The value of $\delta_s$ was obtained by arguing that the singular points in the Euler equation should be canceled by zeros elsewhere. The results were validated using molecular dynamics simulations of hard point particles of different masses with alternate particles having same mass~\cite{chakraborti2022splash}.

In this paper, we generalise the results for the splash problem to an inhomogeneous gas in one dimension. The  system is composed of  vacuum in the region $x < 0$ and an inhomogeneous medium in the region $x \geq 0$, with an initial density distribution $\rho(x) = \rho_0 |x|^{-\beta}$. Using the Euler equation, we determine exactly $\delta_s$ and the shock front growth exponent $\alpha$ as functions of $\beta$. These results are verified using event-driven molecular dynamics (EDMD) simulations. In addition, we find that the thermodynamic quantities obtained from the Euler equation match well with those from the EDMD simulations for all values of $\beta$. 

The remainder of the paper is as follows. In \sref{rescaledEE}, we describe the Euler equation in terms of rescaling functions. In \sref{exaAlp}, we simplify the Euler equation and find the singular points. Using these point we show the curves of $\delta_s$ and singular points with $\beta$. We describe the particles based simulation models for point particles in \sref{EDMD_model}, and show the power law behavior of energy and radius of shock front. In \sref{results}, we compare the similarity exponents and thermodynamic quantities form EDMD and Euler equation. \sref{summary} contains the summary and discussion.

\section{\label{rescaledEE} Model and continuity equations}

In this section, we describe the model, scaling analysis, and the continuity equations describing the evolution of the thermodynamic quantities. Consider a gas in one dimension. Initially the particles are such that it is vacuum in the region $x < 0$ and an inhomogeneous medium in the region $x \ge 0$, with an initial density distribution that varies with  distance $x$ as $\rho(x) = \rho_0 |x|^{-\beta}$, where $0\leq \beta <1$. At time $t = 0$, energy $E_0$ is introduced at $x = 0$. This creates a shock front that propagates into the medium. Let the shock front be at a distance $R(t)$ at time $t$. We assume a power law growth
\begin{equation}
R(t)=A_0 t^\alpha, \label{radius} \quad t \to \infty.
\end{equation}
Due to the vacuum in $x < 0$, the pressure is zero for $x<0$, leading to a reflection of the energy towards the vacuum. At long times, the energy in $x>0$ decreases to zero. We introduce an exponent for this decay:
\begin{equation}
E(t) \sim   t^{-\delta_s}, \quad t \to \infty.
\label{eq:energy_scaling}
\end{equation}
The aim of the paper is to determine $\alpha, \delta_s$ as a function of $\beta$, and then the spatio-temporal behaviour of the different thermodynamic quantities.

The two exponents $\alpha$ and $\delta_s$ are not independent of each other and can be related to each other through scaling arguments. The total number of moving particles in the medium, $N(t)$, scales as  $N(t)\sim R(t)^{1-\beta}$. The typical speed of a particle is $\dot{R}(t)$. Then the total energy of the medium scales as $E(t)\sim N(t)\dot{R}^2 \sim t^{\alpha(3-\beta)-2}$. From Eq.~(\ref{eq:energy_scaling}), we obtain
\begin{align}
\alpha = \frac{2-\delta_s}{3-\beta}, \label{Ealpha}
\end{align}

The propagation of the disturbance into the medium due to the splash at the origin is described by the density $\rho$, velocity $u$, temperature $T$, and pressure $p$ fields. In the hydrodynamic regime, the Euler equation that governs the evolution of these thermodynamic quantities in one dimension are given by~\cite{landaubook,whitham2011linear,barenblatt1996scaling,sedov_book}:
\begin{align}
\partial_t \rho + \partial_x (\rho u) &= 0, \label{massCons} \\
\rho(\partial_t + u \partial_x) u + \partial_x p &= 0, \label{momCons} \\
\partial_t \left( \frac{p}{\rho^3} \right) + u \partial_x \left( \frac{p}{\rho^3} \right) &= 0. \label{eneCons}
\end{align}
By assuming local equilibrium, the local pressure is related to the local temperature and density through the equation of state, which we take to be the ideal gas equation of state:
\begin{align}
p = k_B \rho T, \label{idealEOS}
\end{align}
where $k_B$ is the Boltzmann constant, which we set to $1$.

Across the shock front, the thermodynamic quantities become discontinuous. The values of the thermodynamic quantities at the shock front can be determined by equating the fluxes across the shock front, which leads to the Rankine-Hugoniot boundary conditions~\cite{whitham2011linear,sedov_book,kumar2024shock2}, which adapted to the inhomogeneous gas reduces to
\begin{align}
\rho_1 &= 2 \rho_0 R^{-\beta}, \label{RHmass} \\
u_1 &= \frac{1}{2} \dot{R}, \label{RHvel} \\
p_1 &= \frac{1}{2} \rho_0 \dot{R}^2 R^{-\beta}, \label{RHpres}
\end{align}
where the subscript 1 denotes the quantities just behind the shock front, and $\dot{R}$ is the speed of the shock front.

We define the following non-dimensionalised functions for the different thermodynamic quantities~\cite{barenblatt1996scaling,sedov_book}:
\begin{align}
\xi &= \frac{x}{R(t)}, \label{rescadist}\\
\widetilde{G} &= \frac{R^{\beta}}{\rho_0}\rho(x, t), \label{rescadens} \\
\widetilde{V} &= \frac{u(x, t)}{\dot{R}}, \label{rescavel} \\
\widetilde{Z} &= \frac{3}{\dot{R}^2}  T(x, t). \label{rescatemp} 
\end{align}
Substituting these into the partial differential equations~\eqref{massCons} to~\eqref{eneCons}, we obtain ordinary differential equations for the scaling functions $\widetilde{G}$, $\widetilde{V}$, and $\widetilde{Z}$:
\begin{align}
(\widetilde{V} - \xi) \frac{d \widetilde{G}}{d \xi} + \widetilde{G} \frac{d \widetilde{V}}{d \xi} - \beta \widetilde{G} &= 0, \label{idealmass} \\
(\widetilde{V} - \xi) \frac{d \widetilde{V}}{d \xi} + \frac{1}{3} \frac{d \widetilde{Z}}{d \xi} + \frac{\widetilde{Z}}{3 \widetilde{G}} \frac{d \widetilde{G}}{d \xi} + \left( 1 - \frac{1}{\alpha} \right) \widetilde{V} &= 0, \label{idealmomentum} \\
(\widetilde{V} - \xi) \frac{d}{d \xi} \left( \frac{\widetilde{Z}}{\widetilde{G}^2} \right) + \left( 2 + 2\beta - \frac{2}{\alpha} \right) \frac{\widetilde{Z}}{\widetilde{G}^2} &= 0. \label{idealenergy}
\end{align}
In terms of the scaling functions, the Rankine-Hugoniot boundary conditions~\eqref{RHmass} to~\eqref{RHpres} reduce to
\begin{align}
\widetilde{G}(1) &= 2, \label{idealRHmass} \\
\widetilde{V}(1) &= \frac{1}{2}, \label{idealRHpress} \\
\widetilde{Z}(1) &= \frac{3}{4}. \label{idealRHenergy}
\end{align}

As the splatters recoil into the vacuum, in addition to the Rankine-Hugoniot boundary conditions, we require boundary conditions to smoothly connect the left end of the medium with the right end of the low-density vacuum region. These conditions are~\cite{chakraborti2022splash}:
\begin{align}
\widetilde{G}(\xi \to -\infty) &= 0, \\
\widetilde{V}(\xi \to -\infty) &= -\infty, \\
\widetilde{Z}(\xi \to -\infty) &= 0.
\end{align}

\section{\label{exaAlp} Exact solution of the exponent $\alpha$}

In the splash problem, the energy in the medium decays continuously over time and cannot be determined by scaling arguments, making the problem a self-similar problem of the second kind. Solutions to such self-similar problems have been studied in the context of implosions, where the shock approaches the center of symmetry from infinity~\cite{whitham2011linear,guderley1942powerful,lazarus1981self,hirschler2003self,ponchaut2006imploding}. For self-similar solutions of the second kind, one treats the ordinary differential equations~\eqref{idealmass} to~\eqref{idealenergy} as an eigenvalue problem and seeks the unique value of $\alpha$ for which the solution curves are single-valued over the range of $\xi$. 

After simplifying Eqs.~\eqref{idealmass} to~\eqref{idealenergy} and solving for $d\widetilde{G}/d\xi$, $d\widetilde{V}/d\xi$, and $d\widetilde{Z}/d\xi$, we obtain
\begin{align}
\frac{d\widetilde{G}}{d\xi} &= \frac{\widetilde{G} \left[ (\alpha \beta + \alpha - 1) \left( 3 \widetilde{V}^2 - 2 \widetilde{Z} \right) - 3 \xi \widetilde{V} (2 \alpha \beta + \alpha - 1) + 3 \alpha \beta \xi^2 \right]}{3 \alpha (\widetilde{V} - \xi) \left( (\widetilde{V} - \xi)^2 - \widetilde{Z} \right)}, \label{simpG} \\
\frac{d\widetilde{V}}{d\xi} &= -\frac{\widetilde{Z} [\alpha (\beta - 2) + 2] + 3 (\alpha - 1) \widetilde{V} (\widetilde{V} - \xi)}{3 \alpha \left( (\widetilde{V} - \xi)^2 - \widetilde{Z} \right)}, \label{simpV} \\
\frac{d\widetilde{Z}}{d\xi} &= \frac{2 \widetilde{Z} [3 (\alpha - 1) \xi (\widetilde{V} - \xi) + \widetilde{Z} (\alpha \beta + \alpha - 1)]}{3 \alpha (\widetilde{V} - \xi) \left( (\widetilde{V} - \xi)^2 - \widetilde{Z} \right)}. \label{simpZ}
\end{align}

These equations blow up simultaneously when $\widetilde{Z}(\xi) = (\widetilde{V}(\xi) - \xi)^2$. We denote the singular point as $\xi_s$. In order for the scaling functions $\widetilde{G}$, $\widetilde{Z}$, and $\widetilde{V}$ to be single-valued functions of $\xi$, both the numerator and denominator of Eqs.~\eqref{simpG} to~\eqref{simpZ} must simultaneously vanish at  $\xi = \xi_s$.  The common roots of Eqs.~\eqref{simpG} to~\eqref{simpZ} for which both the numerator and denominator vanish are:
\begin{align}
\widetilde{Z}(\xi_s) &= \frac{9 (\alpha - 1)^2}{(\alpha \beta + \alpha - 1)^2} \xi_s^2, \label{z_singu} \\
\widetilde{V}(\xi_s) &= \frac{\alpha (\beta - 2) + 2}{\alpha \beta + \alpha - 1} \xi_s. \label{v_singu}
\end{align}

\begin{figure}
\centering
\includegraphics[width=\columnwidth]{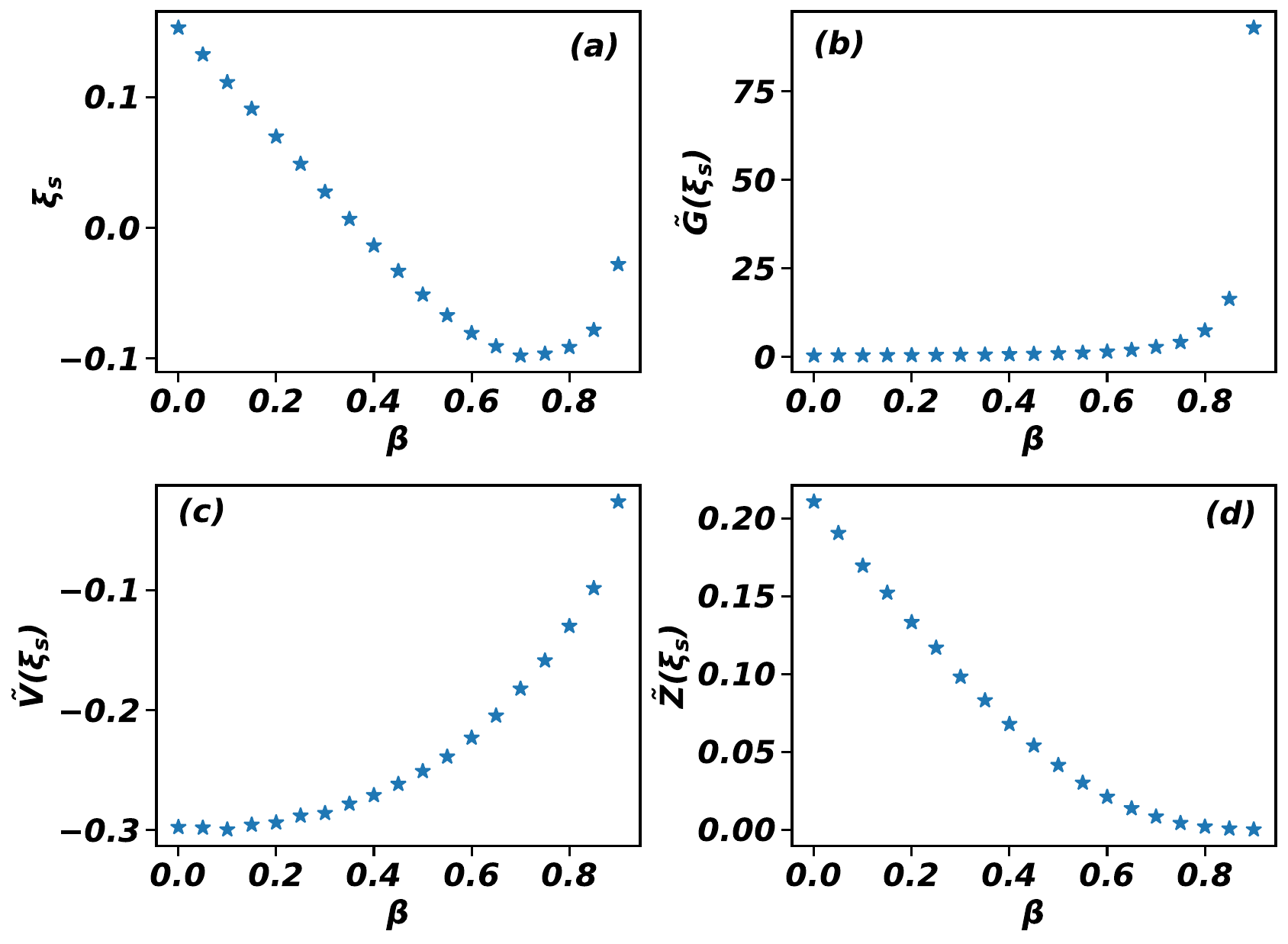}
\caption{The variation with $\beta$ of the different thermodynamic quantities at the singular point $\xi_s$, obtained from the analysis of the Euler equation:  (a) $\xi_s$, (b) density $\widetilde{G}$, (c) velocity $\widetilde{V}$, and (d) temperature $\widetilde{Z}$.}
\label{Tq_sing_pt}
\end{figure}

The above equations allow us to determine, numerically exactly, the exponent $\delta_s$.  The solution curve of the ordinary differential equations~\eqref{simpG} to~\eqref{simpZ} must pass through the singular point \((\widetilde{Z}(\xi_s), \widetilde{V}(\xi_s))\) in the \(\widetilde{Z}\)-\(\widetilde{V}\) plane for the right choice of $\delta_s$. To iteratively determine $\delta_s$, we proceed as follows. We first assign a numerical value to $\delta_s$. We then solve numerically Eqs.~\eqref{simpG}-\eqref{simpZ} using Rankine-Hugoniot boundary conditions. From the solution for $\widetilde{Z}(\xi)$, we determine $\xi_s$ using Eq.~(\ref{z_singu}). This, in turn allows us to determine $\widetilde{V}(\xi_s)$ from Eq.~(\ref{v_singu}). We then check whether the numerical solution of $\widetilde{V}(\xi)$ is consistent with the value of $\widetilde{V}(\xi_s)$. If \(\widetilde{V}(\xi_s)\) from the solution is greater (smaller) than the value obtained from \eref{v_singu}, we increase (decrease) $\delta_s$ and repeat the process, till we obtain $\delta_s$ to desired accuracy. We tabulate the values of $\delta_s$ and $\alpha$, thus obtained, for different values of  $\beta$ in Table~\ref{delta_beta_table}.
In \fref{Tq_sing_pt}, we show the variation of $\xi_s$ and the values of the thermodynamic quantities at the singular points as a function of $\beta$. Curiously, $\xi_s$ is not monotonic with $\beta$. 
\begin{table}
\centering
\caption{\label{delta_beta_table} 
The different values of exponent of energy decay of the inhomogeneous medium for corresponding $\beta$. The different values are obtained by the second-kind self similar solution of the Euler equation (see eqs.~\eqref{simpG} -~\eqref{simpZ}).} 
\begin{tabular}{lll}
\hline
$\beta$ & $\alpha$ & $\delta_s$  \\
\hline
 $0.0$ & $0.627952$ & $0.1161438368$\\
 $0.1$ & $0.653500$ & $0.1048477306$\\
 $0.2$ & $0.681098$ & $0.0929255989$\\ 
 $0.3$ & $0.710967$ & $0.0803888757$\\
 $0.4$ & $0.743350$ & $0.0672880460$\\
 $0.5$ & $0.778503$ & $0.0537414570$\\
 $0.6$ & $0.816672$ & $0.0399860573$\\
 $0.7$ & $0.858056$ & $0.0264699100$\\
 $0.8$ & $0.902714$ & $0.0140277900$\\
\hline
\end{tabular}
\end{table}

We now proceed to compare the values of $\delta_s$ and other thermodynamic quantities obtained from the exact analysis of the Euler equation with those obtained  from molecular dynamics simulations in one dimension. 

\section{\label{EDMD_model} Event driven molecular dynamics simulations} 

We first describe the particle-based model in one dimension that is used for the event-driven molecular dynamics (EDMD) simulations. Consider a line of length $L$, with a set of $N$ point particles labeled $1, 2, \ldots, N$, distributed in the region $L/2 \le x \le L$.  The region $0 \le x < L/2$ is initially empty. To obtain the initial density  $\rho(x) = \rho_0 |x|^{-\beta}$, we assign the position $x_i$ to each particle $i$ as 
\begin{align}
x_i = \frac{L}{2} \left( 1 + q_i^{\frac{1}{1-\beta}} \right),\qquad i=1,2,\hdots,N
\end{align}
where $q_i$ is a random number  drawn from the the uniform distribution on $[0, 1]$.

Initially, all the particles are at rest. At time $t = 0$, a subset of $N_c$ particles, selected from the region near the center, are given initial Gaussian velocities based on their positions:
\begin{align}
u_i = u_0 \exp \left( -\frac{(x_i - L/2)^2}{2\sigma^2} \right),
\end{align}
where $u_0$ and $\sigma$ are positive constants. These initial velocities are then rescaled such that the total initial energy of the system is $E_0$. The system evolves in time through elastic collisions between particles. The particles move ballistically between their successive collisions, and the ordering of the particles is conserved throughout the evolution.

Since head-on collisions of particles with same masses simply exchange their velocities, making the dynamics of the system integrable, in what follows we take these particles as bi-dispersed particles of mass  $m_1$ and  $m_2$ with neighbouring particles having different mass~\cite{du1995breakdown,hurtado2006breakdown,mendl2017shocks}. The post collision velocities $u'_i$ and $u'_j$ of particles $i$ and $j$, having pre collision velocity $u_i$ and $u_j$ respectively, after the collision are
\begin{align}
u_i'=\frac{m_i u_i +  m_j u_j + m_j(u_j-u_i)}{m_i+m_j},\\
u_j'=\frac{m_j u_j +  m_i u_i + m_i(u_i-u_j)}{m_i+m_j}.
\end{align}

The splatter results in particles escaping to the vacuum $x<L/2$.  To avoid finite size effects,  we remove the particles with coordinate  $x < x_0 $, where $0<x_0 \ll L/2$. This is reasonable since the splatters move ballistically towards $x \to -\infty$ and will not interact with the system again.

For our EDMD simulations, we use the following parameters: energy $E_0 = 24$, number of particles $N = 32000$, number of particles excited at the center $N_c = 24$, system size $L = 16000$, and masses $m_1 = 1$ and $m_2 = 2$. We choose the simulation runtime such that the shock front does not reach the boundary at $x = L$.

\section{\label{results}Results}

\subsection{The exponents $\delta_s$ and $\alpha$}

\begin{figure}
\centering
\includegraphics[width=\columnwidth]{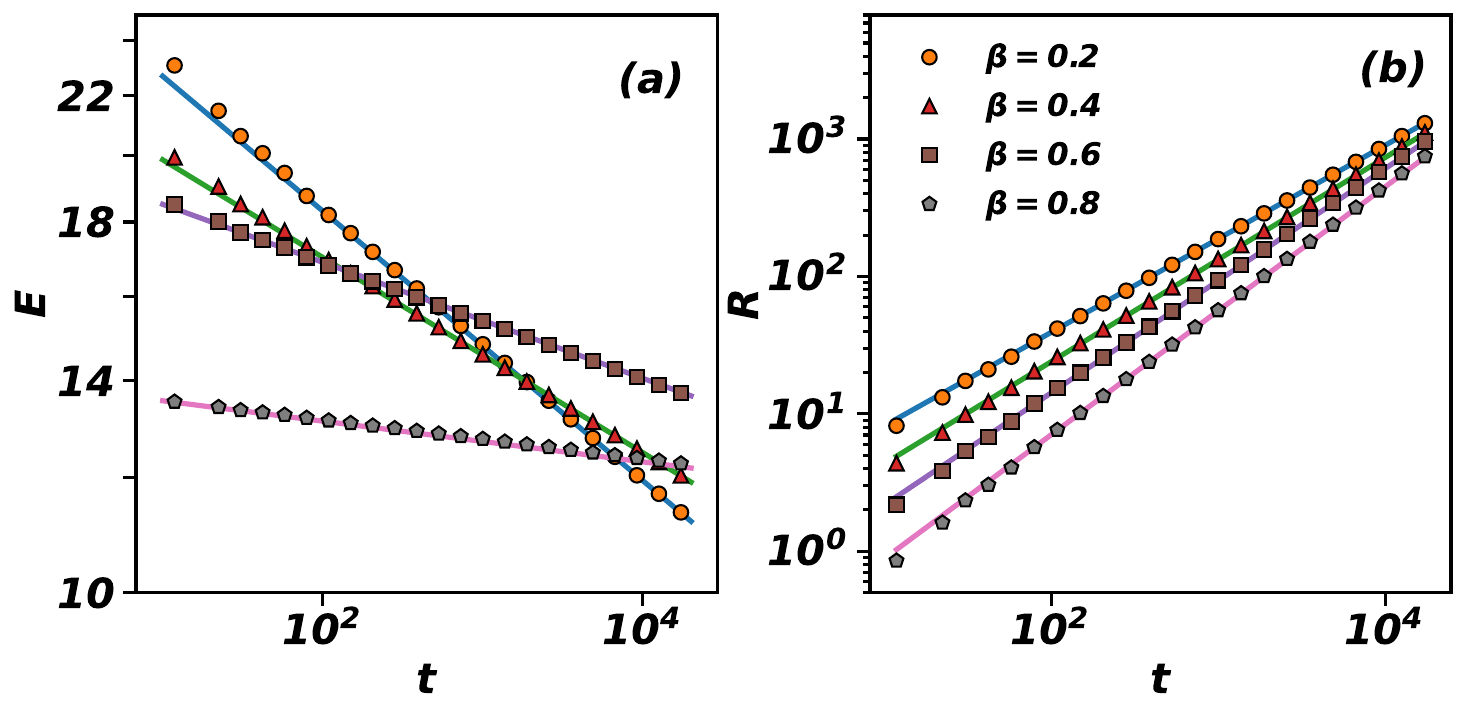}
\caption{ The variation of (a) the energy $E(t)$ of the medium, and (c) the radius of the shock front, $R(t)$, with time $t$ is shown for four different values of $\beta = 0.2$, $0.4$, $0.6$, and $0.8$. The symbols represent the event-driven molecular dynamics data, while the solid lines represent the best fit.}
\label{e_r_t_plot}
\end{figure}

We first show how we extract the exponents $\delta_s$ and $\alpha$ from the EDMD data. The data for energy, obtained as the kinetic energy of all particles with $x>L/2$, and R(t), obtained from the position of the right-most moving particle, shown in \fref{e_r_t_plot}, are power-law in time spanning over more than a couple of decades. We fit the data to power-laws and it can be seen that we obtain excellent fits as the entire data can be fitted to one exponent.

\begin{figure}
\centering
\includegraphics[width=0.666\columnwidth]{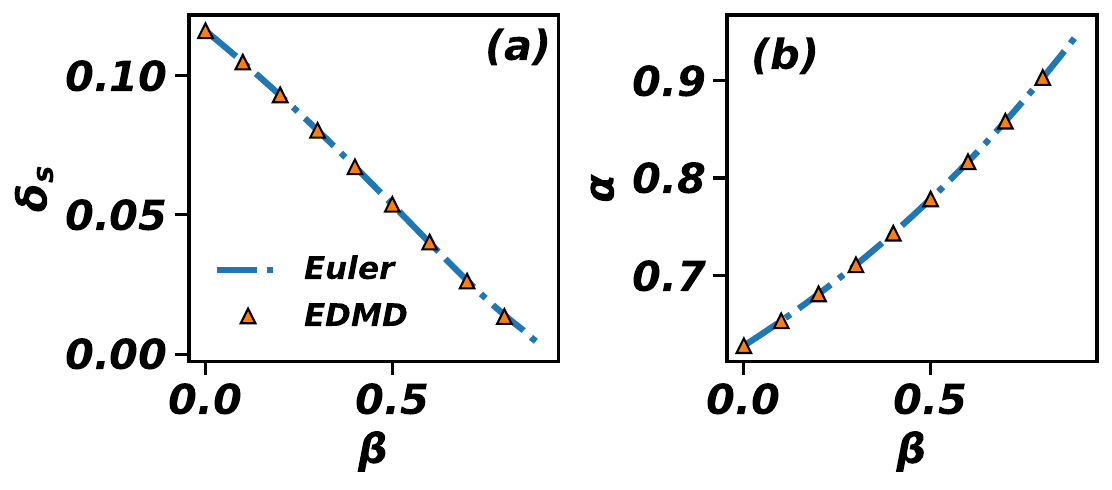}
\caption{ The comparison of the  (a) the energy exponent $\delta_s$ and (b) the shock front radius exponent $\alpha$, obtained from the exact analysis of the Euler equation (dotted lines) with those obtained from the EDMD simulations (symbols). There is excellent agreement. }
\label{delta_alpha_beta}
\end{figure}

The values of  $\delta_s$ and $\alpha$  obtained from the EDMD simulations  are compared with the exponents obtained from the exact analysis of the  Euler equation (see Eqs.~(\ref{simpG}-\ref{simpZ})) in \fref{delta_alpha_beta}. It can be seen that the results obtained from the two analysis are in excellent agreement for the entire range of $\beta$.

\subsection{Behavior of Thermodynamic Quantities}

To measure the thermodynamic quantities $\rho(x,t)$, $u(x,t)$, and $T(x,t)$, we divide the system into bins of size $\Delta$ and compute these quantities at spatial position $x \in [0, L]$. The measured values of these quantities are given by:
\begin{align}
\rho(x,t) &= \left\langle \frac{\sum m_i \delta(x_i, x)}{\Delta} \right\rangle, \label{densityedmd} \\
u(x,t) &= \left\langle \frac{\sum m_i u_i \delta(x_i, x)}{\sum m_i \delta(x_i, x)} \right\rangle, \label{velocityedmd} \\
T(x,t) &= \left\langle \frac{\sum m_i u_i^2 \delta(x_i, x)}{\sum m_i \delta(x_i, x)} \right\rangle - \left( \left\langle \frac{\sum m_i u_i \delta(x_i, x)}{\sum m_i \delta(x_i, x)} \right\rangle \right)^2. \label{tempedmd}
\end{align}
Here, $\langle \cdots \rangle$ denotes the average over different initial configurations, and $\delta(x_i, x)$ is the step function defined as
\begin{align}
\delta(x_i, x) = 
\begin{cases} 
1 & \text{if } |x - x_i| \le \Delta/2, \\
0 & \text{if } |x - x_i| > \Delta/2.
\end{cases}
\end{align}

\begin{figure}
\centering
\includegraphics[width=\columnwidth]{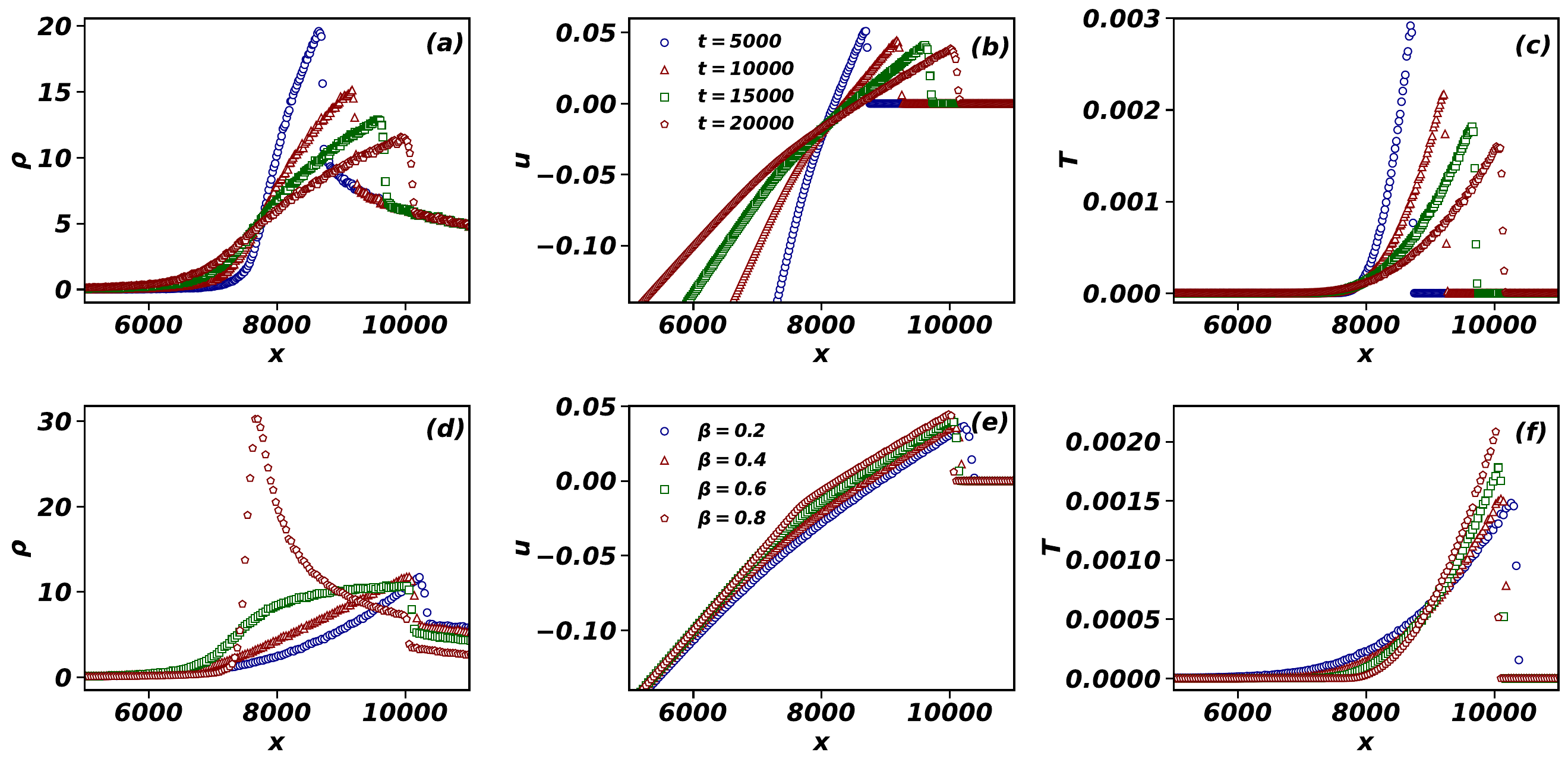}
\caption{The spatial variation of density $\rho(x,t)$, velocity $u(x,t)$, and temperature $T(x,t)$: (a)-(c) fixed $\beta=0.5$ and different times, (d)-(f) fixed time $t=2000$ and different $\beta$. }

\label{thermo}
\end{figure}

In \fref{thermo}, we show the spatial variation of  $\rho(x)$, $u(x)$, and $T(x)$ at four different times for fixed $\beta = 0.5$ [\fref{thermo}(a-c)], and, for fixed time and four values of $\beta$  [\fref{thermo}(d-f)]. We observe that the shock front, indicated by the sharp jumps in the profiles, shifts over time, and the values of thermodynamic quantities at the shock front decrease with time. The results in \fref{thermo}(d--f) show that both the velocity and temperature increase monotonically from the shock center for all values of $\beta$. The density also increases monotonically from the shock center for lower values of $\beta$, reaching a maximum at the shock front. However, for larger values of  $\beta$, the peak in density is no longer at the shock front, and decreases from a peak in the vacuum to the shock front.  

\begin{figure}
\centering
\includegraphics[width=\columnwidth]{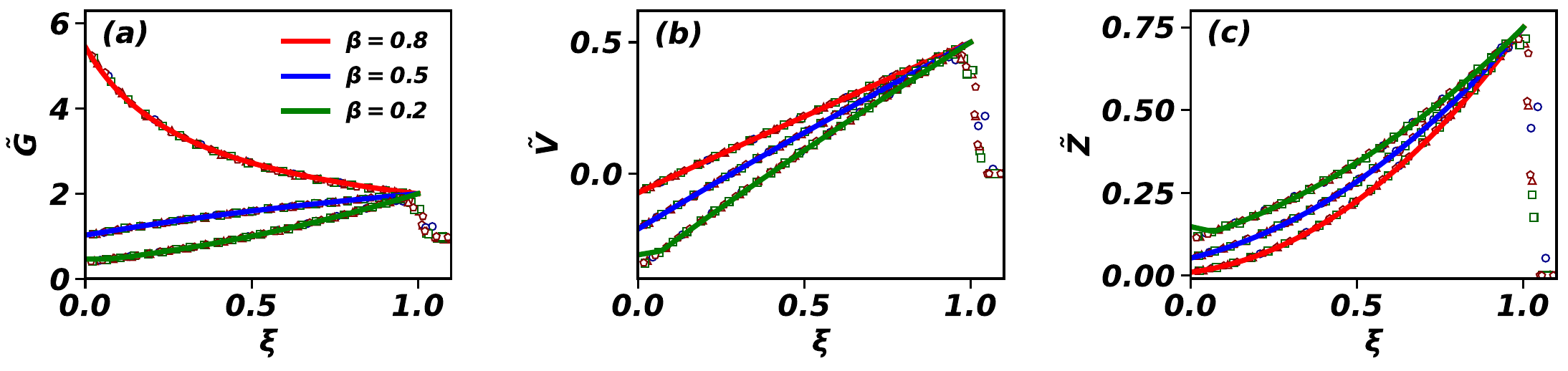}
\caption{The variation of scaling functions: (a) density $\widetilde{G}$, (b) velocity $\widetilde{V}$, and (c) temperature $\widetilde{Z}$ with rescaled distance $\xi$ for $\beta = 0.2$, $0.5$, and $0.8$ is shown.  The symbols represent the EDMD data at four different times, $t = 5000$, $10000$, $15000$, and $20000$, while the solid colored lines represent the self-similar solution of the Euler equation (see Eqs.~(\ref{idealmass}-\ref{idealenergy})). }
\label{thermo_comp}
\end{figure}

Finally, we compare the thermodynamic quantities obtained from the Euler equation (see eqs.~(\ref{idealmass}-\ref{idealenergy})) with those from the EDMD simulations. In  \fref{thermo_comp}(a--c), we see that the non-dimensionalised  thermodynamic quantities at different times collapse onto a single curve, validating the scaling equations~(\ref{rescadist}-\ref{rescatemp}). The scaling function, obtained from the solution of the Euler equation with the exact solution for $\delta_s$ (shown by lines), are in excellent agreement with the EDMD data over the range $\xi \in [0,1]$, for all values of $\beta$ considered. 

\section{\label{summary}Summary and Discussion}

In this study, we investigated the splash problem where energy is initially input into  a system consisting of a vacuum region for \( x < 0 \) and an inhomogeneous medium for \( x \geq 0 \), where the initial density distribution of the medium decays as a power law with radial distance, \( \rho(x) = \rho_0 |x|^{-\beta} \), with \( 0 < \beta < 1 \) in one dimension. At long times, the entire energy is reflected back into the vacuum.  This results in the energy of the inhomogeneous medium decaying over time as a power law, \( E(t) \sim t^{-\delta_s} \).

The  splash problem does not allow for determination of all the exponents using scaling analysis. Instead, we derived the values of \( \delta_s \) and other thermodynamic quantities for different \( \beta \) by solving the Euler equation using self-similarity of the second kind. The solutions were obtained by simplifying the Euler equation's differential terms and identifying the common roots of the numerator and denominator. For the scaling functions to be single-valued in terms of the rescaled distance, the solution curves for different values of \( \beta \) must pass through these common roots.

We validated the numerical values of \( \delta_s \) through event-driven molecular simulations in one dimension. In our simulations, we used bi-disperse particles with masses 1 and 2 units, arranged alternately to prevent the system from becoming integrable, as would occur if all particles had the same mass. The simulation results confirm that the energy of the inhomogeneous medium decays as \( E(t) \sim t^{-\delta_s} \), with \( \delta_s \) matching the value predicted by the Euler equation for the corresponding \( \beta \). Additionally, the radius of the shock front followed the dimensional analysis result, \( R(t) \sim t^{(2-\delta_s)/(3-\beta)} \). We observed a growth in the density peak in the vacuum, which leads to particle confinement between two density peaks: one in the vacuum and the other at the shock front. This results in a much slower decay of the energy, eventually becoming nearly constant. We demonstrated that the thermodynamic quantities derived from the Euler equation closely match the simulation data, and found that the scaling functions in the splash problem exhibit a power-law behavior near the center, akin to the behavior observed in the blast problem in an inhomogeneous medium~\cite{kumar2024shock2}.

The propagation of shocks has also been explored in granular systems, whether induced by a single impact or a continuous source. Notable examples include crater formation due to the impact of high-energy particles on a granular heap~\cite{grasselli2001crater}, the vertical impact of a steel ball into a container of small glass beads~\cite{walsh2003morphology}, or the vertical impingement of gas jets on a granular bed~\cite{metzger2009craters}. Other scenarios involve shock propagation due to the impact of a steel ball on a rapidly flowing granular layer~\cite{boudet2009blast}, the sudden release of localized energy~\cite{jabeen2010universal,pathak2012shock}, or continuous energy injection via the insertion of particles~\cite{joy2017shock}. Granular fingering and pattern formation have also been studied in systems where viscous liquid is injected into dry dense granular material~\cite{cheng2008towards,sandnes2007labyrinth,pinto2007granular,johnsen2006pattern,huang2012granular}. Recently, the \textit{TvNS theory} has been extended to describe shock propagation in granular systems, where energy is not conserved and instead decays continuously over time due to inelastic collisions between particles~\cite{barbier2015blast,barbier2016microscopic}. Studying splash problems in granular systems, where particles cluster due to inelastic collisions, and examining whether energy decays due to splatter at long times could offer an intriguing direction for future research.

\begin{acknowledgements}
The simulations were carried out on the supercomputers Nandadevi and Kamet at The Institute of Mathematical Sciences.
\end{acknowledgements}

\section*{Data availability} The datasets generated during the current study are available from the corresponding author on reasonable request.


\end{document}